\documentclass[twocolumn,showpacs,preprintnumbers,amsmath,amssymb]{revtex4}

\usepackage{graphicx}% Include figure files
\usepackage{dcolumn}% Align columns on decimal point
\usepackage{bm}% bold math

\begin{document}
%\preprint{APS/123-QED}

\title{A Study of the Branching Ratio of $H \rightarrow c\bar{c}$\\
 at a Future $e^{+}e^{-}$ Linear Collider}

\author{Yu, Geum Bong\footnote{Current address: Department of Physics and Astronomy, University of Rochester, Rochester, NY 14627, U.S.A.}}

\author{Kang, JooSang}
\address{Department of Physics, Korea University, Seoul, Korea, 136-701}

\author{Miyamoto, Akiya}
\address{KEK, 1-1 Oho, Tsukuba-Shi, Ibaraki-Ken, Japan, 305-0801}

\author{Park, Hwanbae\footnote{Corresponding author: sunshine@knu.ac.kr}}
\address{Department of Physics, Kyungpook National University, Taegu, Korea, 702-701}

\date{\today}

\begin{abstract}
We carried out a feasibility study on the measurement of the branching ratio 
of $H \rightarrow c\bar{c}$ at a future $e^{+}e^{-}$ linear collider. We used 
topological vertex reconstruction algorithm for accumulating secondary 
vertex information and neural network for optimizing $c$ quark selection. 
With an assumption of a Higgs mass of 120 GeV/$c^{2}$ we estimated 
statistical uncertainty of Br($H \rightarrow c\bar{c}$) to be 20.1\% or 25.7\%
, depending on the number of vertex detector layers, at the center-of-mass 
energy of 250 GeV and the integrated luminosity of 500 ${\rm fb^{-1}}$.
\end{abstract}


\maketitle

\section{\label{sec:level1}Introduction}

In the theory of elementary particles, Higgs boson is introduced to 
elucidate the origin of particle masses. The Standard Model (SM)
proposed the single Higgs doublet that gives rise to a 
scalar particle, while extended supersymmetric models hypothesized that 
two Higgs doublets give separate vacuum expectation values 
to the up-type and down-type quarks. Among the extended models beyond
SM, the Minimal Supersymmetric Standard Model (MSSM) sets the upper 
bound of the lightest neutral Higgs mass around 130 GeV/$c^{2}$\cite{susy}. 
Since the mass of SM Higgs boson is expected to be between 114.4 
GeV/$c^{2}$ and 211 GeV/$c^{2}$ by LEP experiment\cite{lep}, 
the Higgs boson should be identified if it is found. 
Therefore a precise measurement of Higgs couplings to fermions 
at a future $e^{+}e^{-}$ linear collider is indispensable to 
unveil physics of the Higgs sector. In particular, a measurement 
of the branching ratio of $H \rightarrow c\bar{c}$ would provide 
a unique opportunity to study Higgs to up-type quark coupling. 
However, the $H \rightarrow b\bar{b}$ is the dominant process for Higgs 
mass below 140 GeV/$c^{2}$ .

A high efficient charm quark identification is necessary for the 
precise branching ratio study and it is possible only with a high 
performance vertex detector. Development of this vertex detector has been 
a major issue for the linear collider \cite{lcrnd}. To this end, 
understanding detector performance is extremely important 
to physics. 

In this paper, we introduce our method of measuring the branching 
ratio of $H \rightarrow c\bar{c}$ against SM $e^{+}e^{-}$ processes
and other decay branches from Higgs. In addition, we would like to 
see the significance of this precision study with four layers of 
Charge Coupled Device (CCD) and an additional inner layer of 
vertex detector.

\section{\label{sec:level1}HIGGS SIMULATION}

A mass of SM Higgs boson is assumed to be 120 GeV/$c^2$ and
the center-of-mass energy of 250 GeV is selected by the 
highest $s$-channel production cross-section at this energy, 226 fb. 
It is concerned with an integrated luminosity of 500 $\rm fb^{-1}$ for the 
first few years running of the future $e^{+}e^{-}$ linear collider experiment.

Events were generated using PYTHIA 5.7 \cite{pythia} and only the 
Higgsstrahlung process ($e^{+}e^{-} \rightarrow Z^{\ast} 
\rightarrow Z^{\circ}H$) was taken into account.
The branching ratios of the SM Higgs boson was estimated by the 
HDecay program \cite{hdecay}. 
Depending on the decay modes of $Z^{\circ}$, the events are 
categorized into 4-jet mode ($Z^{\circ} \rightarrow q\bar q$
 and $H \rightarrow q\bar q$), 2-jet mode ($Z^{\circ} \rightarrow \nu\bar\nu$ 
 and $H \rightarrow q\bar q$), and charged lepton pair mode ($Z^{\circ} 
\rightarrow \ell^+\ell^-$  and $H \rightarrow q\bar q$). 
Since the branching ratio of Higgs to $c\bar{c}$ is estimated to be very small ($\sim$3\%), 
we concentrated on the 2-jet and 4-jet modes only. 
As background processes from $e^{+}e^{-}$ collision, $W^{+}W^{-},~Z^{\circ}Z^{\circ}$, 
and $q\bar{q}$ events were considered with corresponding cross-sections
 of 15460 fb, 1250 fb, and 47300 fb, respectively.

The detector simulation was performed using a fast parameterized simulator \cite{jlc}, which is 
implemented in the Joint Linear Collider (currently called as Global Linear 
Collider) detector \cite{report}. In this simulator five parameters 
of the helical track and their error matrices including non-diagonal elements 
were generated; thus the quality of the vertices is similar to that of a 
full simulation. The JLC vertex detector is equipped with four layers of 
CCD at the radius from 2.4 cm to 6 cm of which 
intrinsic spatial resolutions are 4 ${\rm \mu m}$ in ${\rm r \phi}$ and z 
directions. The solenoidal magnet field of the detector is 3 tesla. With 
a vertex detector constraint, the impact parameter resolution in xy plane
(${\rm \sigma_{r \phi}}$) is ${\rm \sqrt{(25/p sin^{2/3} \theta)^{2} +4^2}
~\mu m}$, and the momentum resolution for charged track
(${\rm \Delta p_t/p_t }$) is ${\rm \sqrt{(1\times 10^{-4} p_t)^{2}+ 10^{-3} } }$. 
This study was done with four/five CCD layers of vertex detector parameter, 
repectively, to see the influence of the vertex detector options on 
the physics result. The vertex detector parameter sets are shown in the 
Table~\ref{tab:table1}.

\begin{table}
\begin{ruledtabular}
\begin{tabular}{lcrcrcrcrc}
number of CCD layers & Beam pipe & 1st & 2nd & 3rd & 4th & 5th \\
\hline
4 & 2.0 & 2.4 & 3.6 & 4.8 & 6.0 & \\
5 & 1.0 & 1.2 & 2.4 & 3.6 & 4.8 & 6.0  \\
\end{tabular}
\caption{\label{tab:table1} Two vertex detector parameter set. 
the number of CCD layers and their positions(in cm) from the center of 
the beam pipe, the radius of beam pipe are listed for each set. }
\end{ruledtabular}
\end{table}

In the event reconstruction, jets were reconstructed using JADE clustering 
algorithm~\cite{jade} and the vertices were reconstructed using the topological 
vertex reconstruction algorithm (ZVTOP program)~\cite{zvtop}. Based on kinematic 
variables and reconstructed vertex information, an artificial neural network (NN)
 \cite{nn} is formed to identify $H \rightarrow c\bar{c}$ events more efficiently.  

We considered the non-$c\bar{c}$ Higgs decays ($H \rightarrow b\bar{b},~gg,~
WW^{\ast}$; Higgs background) and other processes from $e^{+}e^{-}$ collision
 ($e^{+}e^{-} \rightarrow Z^{\circ}Z^{\circ},~W^{+}W^{-},~q\bar{q}$; non-Higgs 
background) as backgrounds of $H \rightarrow c\bar{c}$ measurement.

\section{\label{sec:level1}Selection of $H \rightarrow c\bar{c}$}

\subsection{\label{sec:level2}2-jet mode}

\begin{figure}[h]
\resizebox{9cm}{!}{\includegraphics{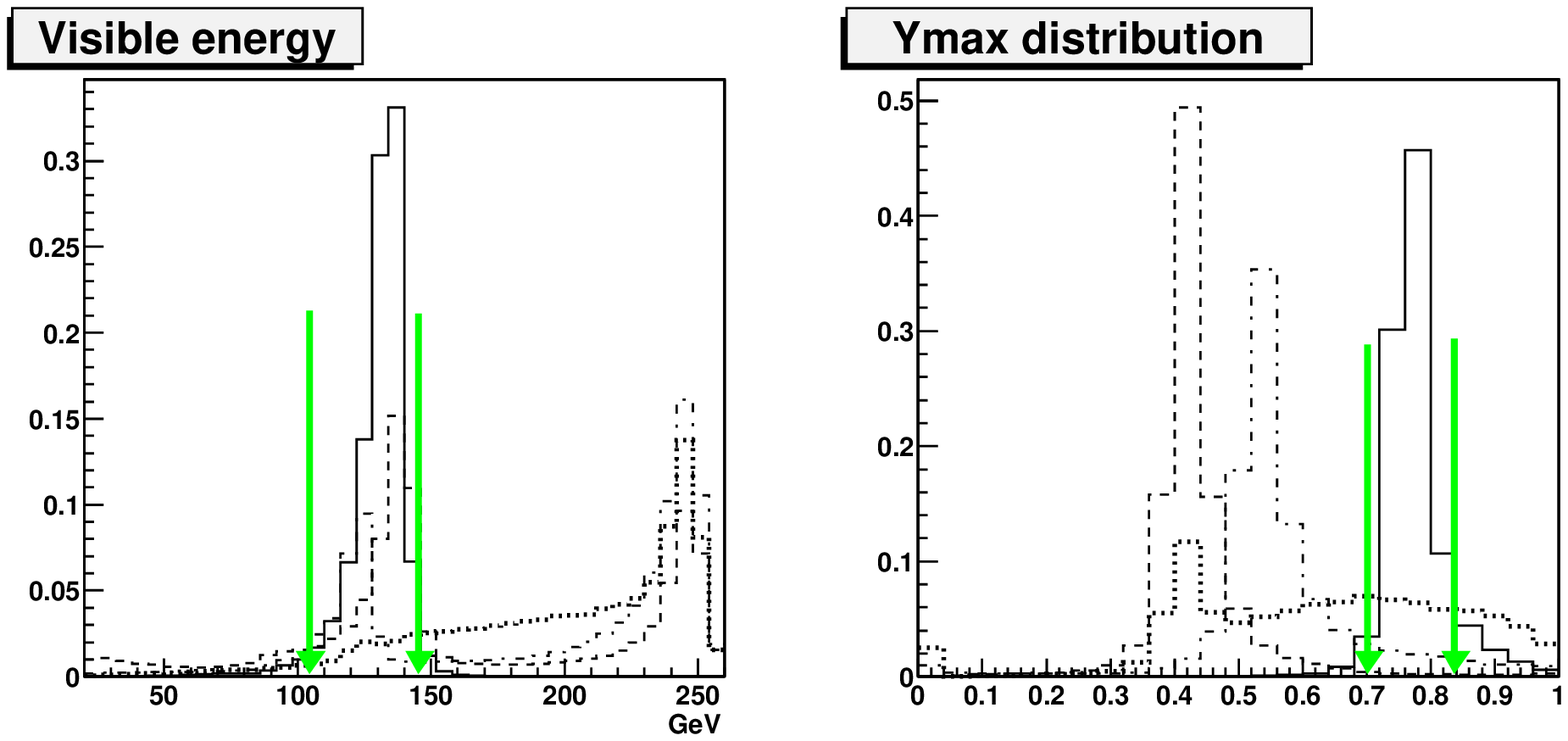}}
\caption{\label{fig:2jsel} Visible energy and Ymax distributions in 2-jet mode are shown.
Arrows are place on the cut value. Solid: $Z^{\circ}H$, dotted: $W^{+}W^{-}$, dash-dotted: 
$Z^{\circ}Z^{\circ}$, dashed: $q\bar{q}$. }
\end{figure}

In 2-jet mode all clustered particles are forced to make 2 jets by 
adjusting the maximum value of ycut~\footnote{${\rm y_{ij}=\frac{2E_{i}E_{j}
(1-cos\theta_{ij})}{E_{visible}^2}}$, where ${\rm E_i~and~E_j}$ are energy 
of i-th and j-th cluster, respectively, and $\theta_{ij}$ is angle 
between two clusters.}, Ymax. The following selections are applied 
to reduce non-Higgs background in the sample: (1)visible energy between 
110 GeV and 143 GeV, (2)missing transverse momentum between 25 GeV/$c$ 
and 70 GeV/$c$, (3)Higgs mass between 105 GeV/$c^{2}$ and 125 GeV/$c^{2}$, 
(4)recoiled $Z^{\circ}$ mass between 82 GeV/$c^{2}$ and 120 GeV/$c^{2}$, 
(5)thrust between 0.75 and 0.99, (6)Ymax between 0.72 and 0.84, and 
(7)mass of each jet between 2 GeV/$c^{2}$  and 40 GeV/$c^{2}$.
 Here the Higgs mass is an invariant mass of all the observed particles
 and the recoiled $Z^{\circ}$ mass is calculated disregarding the 
initial state radiation. 
We also require a successful secondary vertex reconstruction and at least one 
${\rm P_{t}}$ corrected invariant mass (MSPTM, ${\rm \sqrt{ {M^2}_{VTX}+{P_t}^{2}
 }+|P_{t}|}$) of the secondary vertex to be between 0.1 GeV/$c^{2}$ and 7.0 
GeV/$c^{2}$, where ${\rm P_t}$ is the total transverse momentum of the 
secondary tracks with respect to the flight direction of the vertex.
The separation between signal and non-Higgs background is clearly seen 
in Fig.~\ref{fig:2jsel}, while Higgs background is not distinguishable in this aspect.  
   
After the sample selection, NN is trained to maximize the 
selection quality. Two sets of training are used; one against non-Higgs 
backgrounds (Background NN-training) and the other against Higgs backgrounds (
Higgs NN-training). For each NN-training $H \rightarrow WW^{\ast}$ and 
$e^{+}e^{-} \rightarrow q\bar{q}$ processes are not used due to the 
featureless patterns and small statistics, respectively. \\

\begin{figure}[h]
\resizebox{9cm}{!}{\includegraphics{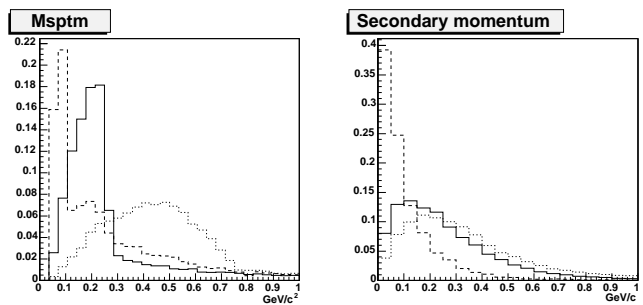}}
\caption{\label{fig:2h} MSPTM and the corrected secondary momentum distributions
are used as input patterns for $H \rightarrow c\bar{c}$ distinction from 
Higgs backgrounds in 2-jet mode. Solid: $H\rightarrow c\bar{c}$, 
dotted: $b\bar{b}$, dashed: $gg$.}
\end{figure}

\begin{figure}[h]
\resizebox{9cm}{!}{\includegraphics{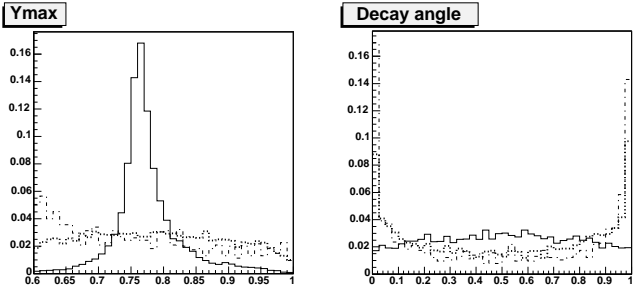}}
\caption{\label{fig:2b} Ymax and decay angle distributions show good separation
 between $c\bar{c}$ and non-Higgs backgrounds in 2-jet mode. They are used as 
 input patterns for Background NN-training. Solid:$Z^{\circ}H$, dotted:$W^{+}W^{-}$, 
dash-dotted:$Z^{\circ}Z^{\circ}$.}
\end{figure}

The normalized input patterns for NN-trainings were given from the ZVTOP program;
the number of vertices, MSPTM, transverse momentum of the secondary vertex, 
decay length, invariant mass of the secondary vertex, number of tracks in 
the secondary vertex, and momentum of the secondary vertex divided by total 
momentum (corrected secondary momentum). Distinctive input patterns for Higgs 
NN-training and Background NN-training are shown in Fig. \ref{fig:2h} and 
Fig. \ref{fig:2b}, respectively. In the Background NN-training
Ymax and jet decay angle were included in the input patterns especially 
for this 2-jet mode.

\begin{figure}[h]
\resizebox{9cm}{!}{\includegraphics{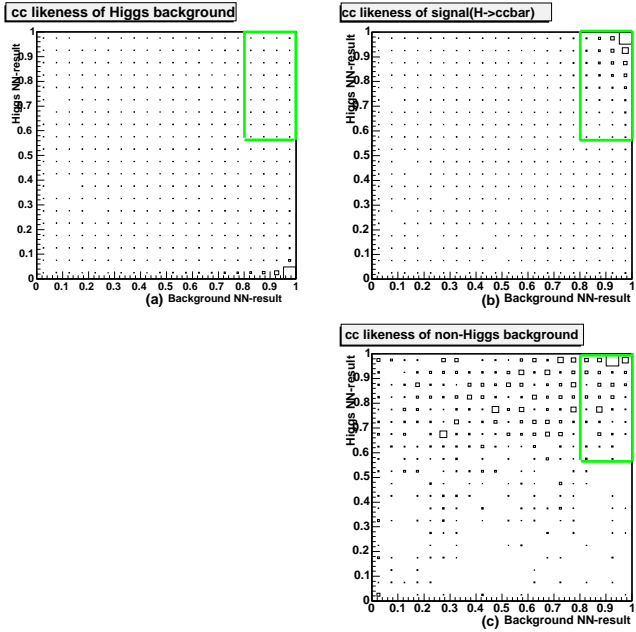}}
\caption{\label{fig:2d2j} 2D NN-result for $H \rightarrow c\bar{c}$ selection
in 2-jet mode;  (a)Higgs backgrounds, 
(b)$H \rightarrow c\bar{c}$ signal event, and (c)non-Higgs backgrounds.
The signal region is seen at the top right. }  
\end{figure}

Using both results of Higgs NN-training and Background NN-training,  
the signal region is evaluated in two dimensional plane for best significance. 
The signal selection area is shown in the top right corner of the Fig.~\ref{fig:2d2j}.

%
% 2d signal fig -> 2j_3t_signal1.eps 
%

\subsection{\label{sec:level2}4-jet mode}

\begin{figure}[h]
\resizebox{9cm}{!}{\includegraphics{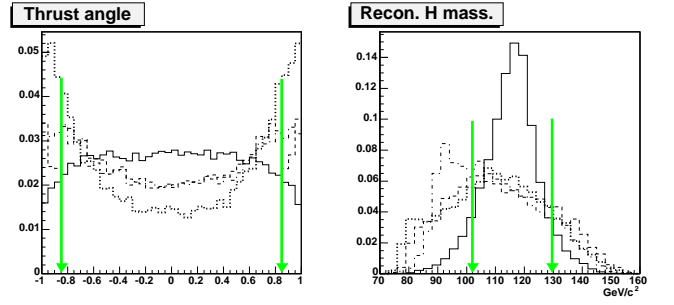}}
\caption{\label{fig:4j} Thrust angle and Higgs mass distributions in 4-jet mode
are shown. Arrows are placed on the cut value. Solid: $Z^{\circ}H$, 
dotted: $W^{+}W^{-}$, dash-dotted: $Z^{\circ}Z^{\circ}$, dashed:
$q\bar{q}$. }
\end{figure}
%
%  4j-selection fig. -> 4j_sel1.eps
%

\begin{figure}[h]
\resizebox{9cm}{!}{\includegraphics{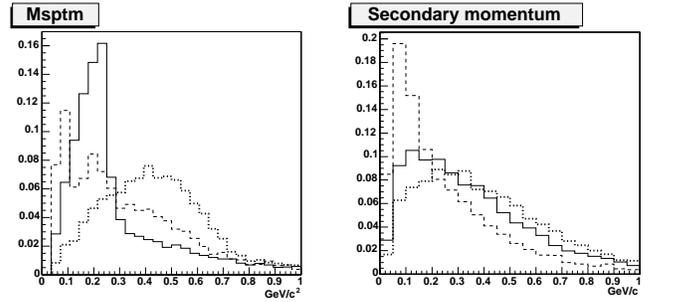}}
\caption{\label{fig:4h}MSPTM and the corrected secondary momentum 
distributions in 4-jet mode. They are used for $H \rightarrow c\bar{c}$ distinction from 
non-$c \bar{c}$  Higgs decays in Higgs NN-training. Solid: $H \rightarrow c\bar{c}$, 
dotted: $b\bar{b}$, dashed: $gg$.}
\end{figure}

\begin{figure}[h]
\resizebox{9cm}{!}{\includegraphics{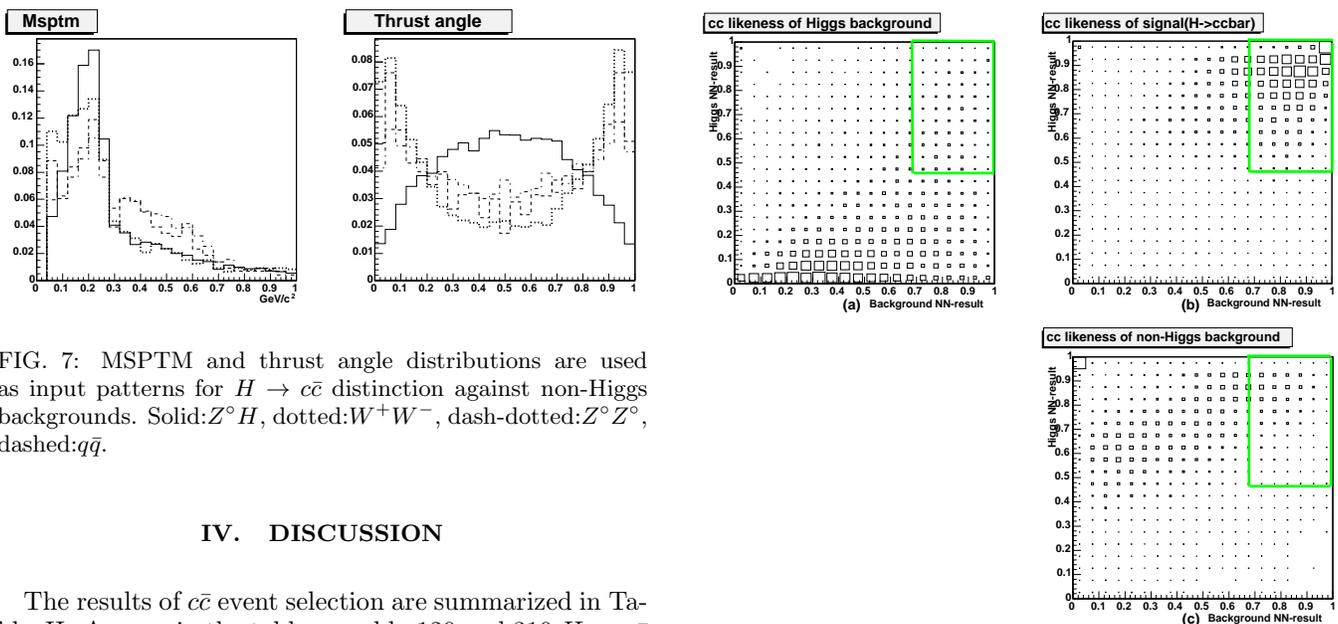}}
\caption{\label{fig:4b}MSPTM and thrust angle distributions are used as 
input patterns for $H \rightarrow c\bar{c}$ distinction against non-Higgs backgrounds. 
Solid:$Z^{\circ}H$, dotted:$W^{+}W^{-}$, dash-dotted:$Z^{\circ}Z^{\circ}$, 
dashed:$q\bar{q}$.}
\end{figure}

%
% 4j background train. -> 4bq_train.eps 
%

%
% 4j hig train. -> 4h_train.eps
%

%
% 4j 2d-signal fig. -> 4j_3t_signal.eps 
%

\begin{figure}
%\resizebox{9cm}{!}{\includegraphics{./fig/4j_3t_signal3}}
\resizebox{9cm}{!}{\includegraphics{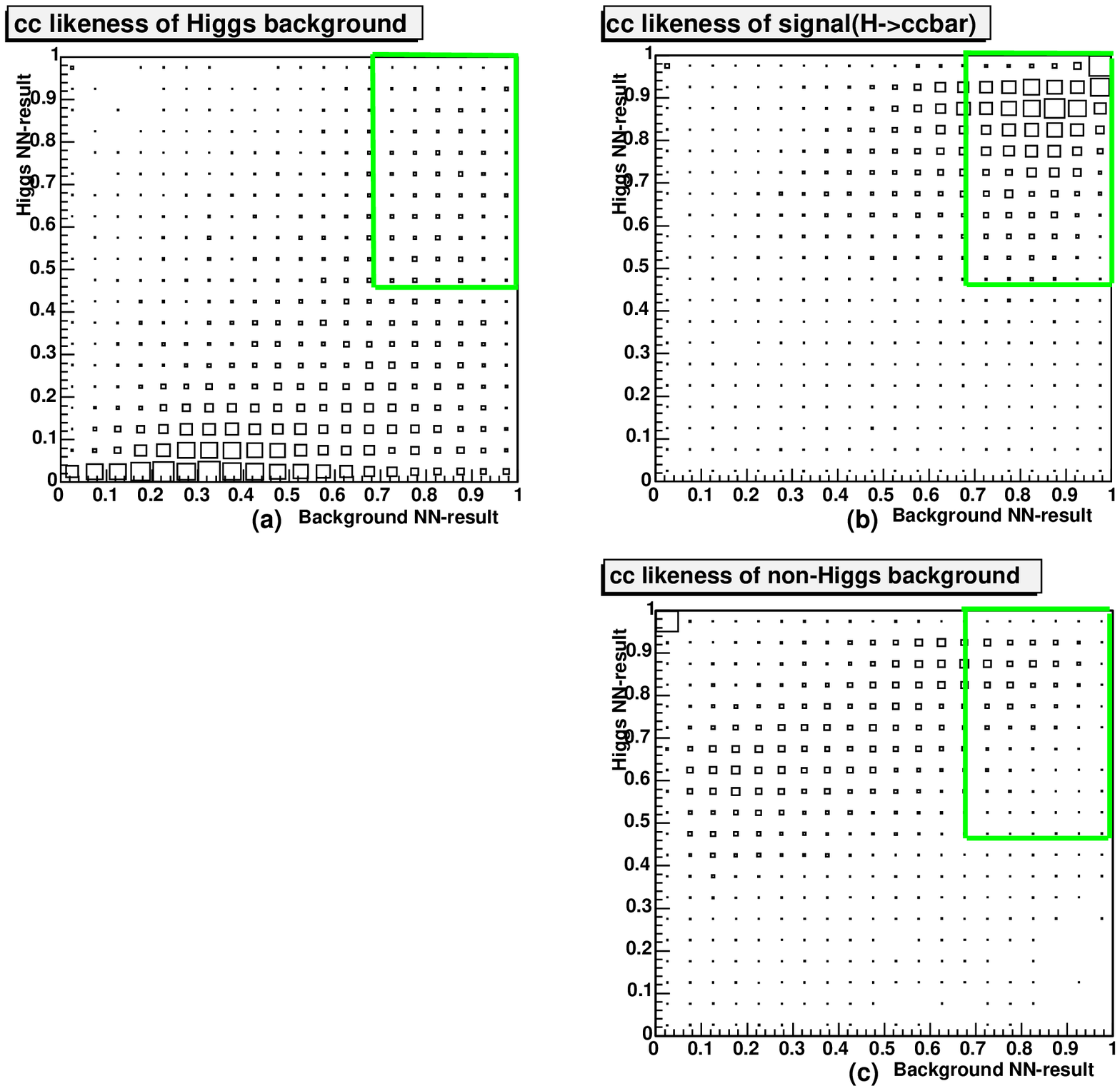}}
\caption{\label{fig:2d4j}2D NN-results for $H \rightarrow c\bar{c}$ selection 
in the 4-jet mode;  (a)Higgs backgrounds, 
(b)$H \rightarrow c\bar{c}$ signal event, and (c)non-Higgs backgrounds. 
The signal region is seen at the top right.}
\end{figure}

Here we used forced 4-jet clustering for Higgs jet pair and $Z^{\circ}$ jet pair.
The jets are identified by minimum $\chi^{2}$, which is defined as the squared sum of the 
differences between reconstructed invariant mass of jet pair and the expected 
mass divided by each mass resolution: 
\begin{eqnarray*}
\chi^2 = (\frac{M_{Z^{\circ}}-91.2}{\rm width})^2 + 
(\frac{M_{{\rm recoil}Z^{\circ}}-91.2}{\rm width})^2 \\
+ (\frac{M_H-120}{\rm width})^2 + (\frac{M_{{\rm recoil} H}-120}{\rm width})^2.  
\end{eqnarray*}

The selection cuts were made: (1) visible energy greater than 210 GeV, (2) 
$\cos \theta$ of thrust axis(thrust angle) between $-0.85$ and 0.85, 
(3) Higgs mass between 103 GeV/$c^{2}$ and 130 GeV/$c^{2}$, (4) $Z^{\circ}$
 mass between 78 GeV/$c^{2}$ and 102 GeV/$c^{2}$, (5) mass recoiled to the Higgs 
jet pair greater than 80 GeV/$c^{2}$, (6) number of particles in each jet of 
Higgs pair greater than 5, and (7) Ymax value greater than 0.01.

Only 2 jets which are assigned to Higgs are used for the vertex tagging.
We also required at least one successful secondary vertex 
reconstruction with MSPTM between 0.1 GeV/$c^{2}$ and 7.0 GeV/$c^{2}$ out of 2 
jets from Higgs. The distributions of Higgs mass and 
thrust angle are shown in the Fig.~\ref{fig:4j}. With the same 
procedure as the 2-jet study, we trained the NN against non-Higgs background and
Higgs background separately using normalized vertex information from the ZVTOP program. 
The MSPTM and the corrected secondary momentum 
distributions of the $c\bar{c}$ and the non-$c \bar{c}$ Higgs decays are seen in the 
Fig. \ref{fig:4h}. We did not use $H \rightarrow WW^{\ast}$ events 
in Higgs NN-training due to their featureless patterns.
 In the Background NN-training, thrust angle is added in the input pattern. 
As seen in the Fig. \ref{fig:4b}, the thrust angle shows a clear 
separation between the signal and non-Higgs backgrounds. 
We use both results of Higgs NN-training and Background NN-training same as 2-jet mode.
The 2D NN-results are shown in the Fig. \ref{fig:2d4j} with signal box at the top right.

\begin{table*}
\begin{ruledtabular}
\begin{tabular}{ccccccc}
Number of CCD layers & \multicolumn{2}{c}{4 CCD layers} & 
\multicolumn{2}{c}{5 CCD layers}\\
\hline
Decay mode  $\backslash$ Events & 2-jet & 4-jet & 2-jet & 4-jet \\
\hline
${H \rightarrow c\bar{c}}$ & 122.7(17.6) & 306.3(12.6) & 112.7(16.2) &
 316.8(13.0) \\
${H \rightarrow b\bar{b}}$ & 641.8(4.2) & 2686.7(5.0) & 231.2(1.5) & 
1807.2(3.4) \\
${H \rightarrow gg}$ & 28.1(1.8) & 217.6(3.9) & 8.9(0.6) & 143.4(2.6)  \\
${H \rightarrow WW^{\ast}}$ & 23.6(0.8) & 226.7(2.1) & 10.1(0.3) &
 178.0(1.7) \\
\hline
${e^{+}e^{-} \rightarrow W^{+}W^{-}}$ & 640(0.008) & 9790(0.130) & 330(0.004) &
 8530(0.110) \\
${e^{+}e^{-} \rightarrow Z^{\circ}Z^{\circ}}$ & 100(0.016) & 1710(0.274) &
 30(0.004) & 1305(0.209)  \\
${e^{+}e^{-} \rightarrow q\bar{q}}$ & 
20($<$0.001) & 1730(0.007) & 5($<$0.001) & 1345(0.006)\\
\hline
S/N & 0.0843 & 0.0187 & 0.1832 & 0.0238\\
${S / \sqrt{S+N}}$ & 3.09 & 2.37 & 4.17 & 2.71 \\
Statistical uncertainty & \multicolumn{2}{c}{25.7 $\%$} & 
\multicolumn{2}{c}{20.1 $\%$}   \\
\end{tabular}
\caption{\label{tab:table2}The number of $c$ quark tagged events, efficiency
(percentage in parenthesis), and significance (last three rows) of ${ H 
\rightarrow c\bar{c}}$ measurements for two different vertex detector 
parameters. The statistical uncertainty (last low) is calculated combining 
significance of 2-jet and 4-jet. We used 100 ${\rm fb^{-1}}$ in this 
analysis and scaled up to 500 ${\rm fb^{-1}}$ in this table.}
\end{ruledtabular}
\end{table*}

\section{\label{sec:level1}Discussion}

The results of $c\bar{c}$ event selection are summarized 
in Table ~\ref{tab:table2}. As seen in the table, roughly 120 and 310 
$H \rightarrow c\bar{c}$ events are selected in 2-jet and 4-jet modes, 
respectively, with a reasonable significance despite its small branching 
ratio of $H \rightarrow c\bar{c}$ and large backgrounds from non-Higgs processes 
and non-$c \bar{c}$ Higgs decays. 

In the $H \rightarrow c\bar{c}$ measurement in 2-jet mode, major backgrounds are 
those from $H \rightarrow b\bar{b}$ where $b$ is misidentified as $c$ and 
$W^{+}W^{-}$ events which are reconstructed as 2-jet events due to imperfect 
detector acceptance. $e^{+}e^{-} \rightarrow e \nu W$ is not considered here 
due to relatively small cross-section than other background processes.

In the case of 4-jet mode, $W^{+}W^{-}$ and $Z^{\circ}Z^{\circ}$ events
are increased after event selection because two $c$ jets can be produced in 
the final state of these processes. In addition to the c-jet contamination 
from backgrounds, there is additional ambiguity in selecting two Higgs jets 
out of four jets. Combining these effects, the 
significance in the 4-jet mode is less than 3$\sigma$ and worse than the 2-jet mode. The 
statistical uncertainty combining 2-jet and 4-jet analyses is estimated as 25.7\% 
in the case of four CCD layers of the vertex detector. Similar analysis with different detector
configuration was report\cite{JB}
for assumption of the same Higgs mass and luminosity at $\sqrt s$=350 GeV and 500 GeV  
that the relative braching ratio errors be 19\% and 39\%, respectively.

When an additional CCD layer of the vertex detector is included near the
interaction point, the impact parameter resolution for low momentum track is 
improved since the lever arm for track extrapolation to the interaction point 
is reduced. Thus we can reconstruct secondary vertices much closer to the 
interaction point. 
The better separation of $b$- and $c$- jets reduces backgrounds in the event selection, 
especially in 2-jet mode.
 Improvements in 4-jet selection is not decisive, suffering ambiguities in 
Higgs jet selections. Combining 2-jet and 4-jet, relatively 20\% improvement
 in background reduction is achieved compared to the study with four CCD layers of the vertex
 detector. 

\section{\label{sec:level1}Summary}

In this study we focused on the $H \rightarrow c\bar{c}$ measurements in 
2-jet and 4-jet modes in the case of 120 GeV/$c^{2}$ Higgs mass at the
center-of-mass energy of 250 GeV in the future $e^{+}e^{-}$ linear collider. 
In the study, the topological vertex finding algorithm
 was used for tagging the $c$-jet and the neural network was used to 
optimize the $H \rightarrow c\bar{c}$ selection. 

With consideration of 500 ${\rm fb^{-1}}$ data, we obtained the statistical 
uncertainty of 25.7\% for the measurement of $H \rightarrow c\bar{c}$ 
with four CCD layers vertex detector. The statistical uncertainty would be 
improved to 20.1\% with an additional CCD layer of vertex detector 
near the interaction point. 

\section{\label{sec:ack}Acknowledgment}

We would like to thank the members of ACFA Joint Linear Collider Physics and 
Detector Working group \cite{acfa} for valuable
 discussions during the course of the analysis. We also thank Dr. David 
Jackson for allowing us to use his ZVTOP program. 
We acknowledge the support by Korea Research Foundation Grant funded by Korea 
Government (MOEHRD) (KRF-2005-070-C000321) and Korea Science and Engineering Foundation. 
Akiya Miyamoto is partially 
supported by Japan-Europe Research Cooperative Program.

\end{document}